\documentclass[aps,prx, amsfonts, superscriptaddress, amssymb, amsmath, reprint, showkeys, nofootinbib, raggedbottom, longbibliography]{revtex4-1}
\usepackage[english]{babel}
\usepackage[utf8]{inputenc}
\usepackage[colorinlistoftodos, color=green!40, prependcaption]{todonotes}
\usepackage{amsthm}
\usepackage{mathtools}
\usepackage{physics}
\usepackage{xcolor}
\usepackage{graphicx}
\usepackage{adjustbox}
\usepackage{placeins}
\usepackage[T1]{fontenc}
\usepackage{lipsum}
\usepackage{csquotes}
\usepackage{float}
\usepackage{afterpage}
\usepackage{bm}
\usepackage{booktabs}
\usepackage[version=4]{mhchem}

\newcommand{\phasesect}{1}
\newcommand{\moresans}{4}
\newcommand{\phaseevofig}{6}
\newcommand{\phasechoice}{9}

\newcommand{\IVKas}{11}

\newcommand{\VIIIKas}{16}
\newcommand{\VIIIKdas}{17}

\newcommand{\CCCmTdas}{22}
\newcommand{\IIKa}{23}
\newcommand{\VKa}{24}

\newcommand{\VIIIKa}{26}

\usepackage[normalem]{ulem}

\usepackage[pdftex, pdftitle={Article}, pdfauthor={Author}]{hyperref} % For hyperlinks in the PDF
\graphicspath{ {Figures/} }

\begin{document}
\title{Cascade of Spin Moir\'e Superlattices with In-Plane Field in Triangle Lattice Semimetal \ce{EuAg4Sb2}}

\author{Paul M. Neves}
\email{pmneves@mit.edu}
\affiliation{Department of Physics, Massachusetts Institute of Technology, Cambridge, MA 02139, USA}
\author{Takashi Kurumaji}
\affiliation{Division of Physics, Mathematics and Astronomy, California Institute of Technology, Pasadena, California 91125, USA}
\author{Joshua P. Wakefield}
\affiliation{Department of Physics, Massachusetts Institute of Technology, Cambridge, MA 02139, USA}
\author{Chi Ian Jess Ip}
\affiliation{Department of Physics, Massachusetts Institute of Technology, Cambridge, MA 02139, USA}
\author{Robert Cubitt}
\affiliation{Institut Laue Langevin, 71 Avenue des Martyrs, F-38000 Grenoble Cedex 9, France}
\author{Satoru Hayami}
\affiliation{Graduate School of Science, Hokkaido University, Sapporo 060-0810, Japan}
\author{Jonathan S. White}
\affiliation{PSI Center for Neutron and Muon Sciences, CH-5232 Villigen PSI, Switzerland.}
\author{Joseph G. Checkelsky}
\affiliation{Department of Physics, Massachusetts Institute of Technology, Cambridge, MA 02139, USA}

\date{\today} % Leave empty to omit a date


\maketitle
\section{Abstract}
\ce{EuAg4Sb2} is a rhombohedral europium triangle lattice material that exhibits a rich phase diagram of spin moir\'{e} superlattices (SMS) and single-$q$ magnetic phases. In this paper, we characterize the incommensurate phases accessible with field applied in the plane with small angle neutron scattering (SANS). A variety of phases with unusual SANS patterns are accessible with magnetic field applied along the $a$ and $a^*$ directions. Many of these phases can be understood to be multi-$q$ phases. One phase in particular, ICM2b (ICM=incommensurate magnetic phase), is rather unconventional in that it is an anisotropic multi-$q$ phase that can rotate freely within the $ab$-plane, dependent on magnetic field direction and history. The stabilization of tunable multi-$q$ incommensurate spin textures \textit{via} in-plane field sets this class of materials apart from conventional skyrmion materials. We further identify that the propagation vectors of the in-plane phases have a significant commensuration with the diameter of the smallest pocket of the Fermi surface ($2k_{\text{F}}$). The multi/single-$q$ nature is also correlated with the enhancement of resistivity, suggesting that a gap opens in the electron bands at $q=2k_{\text{F}}$. We also compare with a phenomenological model of the phase diagram. The richness of phases revealed in this study hint at the frustrated nature of the incommensurate magnetism present in \ce{EuAg4Sb2} and motivate further probes of these phases and the origin of the stability of spin moir\'{e} superlattices. Finally, the coupling of the multi-$q$ nature and $q=2k_{\text{F}}$ commensuration condition reveals the key requirements for a strong SMS transport response.

% 150 words
% \ce{EuAg4Sb2} is a rhombohedral europium triangle lattice material that exhibits a rich phase diagram of spin moir\'{e} superlattices and single-$q$ magnetic phases. In this paper, we characterize a variety of unusual phases accessible with field applied in the plane with small angle neutron scattering (SANS). One phase in particular, ICM2b, is rather unconventional in that it is an anisotropic multi-$q$ phase that can rotate freely within the $ab$-plane. We further identify that the propagation vectors of the in-plane phases have a significant commensuration with the diameter of the smallest pocket of the Fermi surface ($2k_{\text{F}}$). The multi/single-$q$ nature is also correlated with the enhancement of resistivity, suggesting that a gap opens in the electron bands at $q=2k_{\text{F}}$. We also compare with a phenomenological model of the phase diagram. The observed coupling of the multi-$q$ nature and $q=2k_{\text{F}}$ commensuration condition reveals the key requirements for a strong SMS transport response.

\section{Introduction}
Incommensurate spin modulation instabilities in metals are often understood as being driven by an energy gain through the opening of Fermi surface gaps \cite{solenov2012chirality}. Density wave instabilities are also often present in the phase diagrams of a variety of quantum materials \cite{stewart2011superconductivity, manzeli20172d, hayden2024charge}, making their exploration highly beneficial to understanding coupled degrees of freedoms in solids. Recently, a variety of incommensurate multi-$q$ spin textures (meaning magnetic phases formed from a superposition of multiple magnetic modulations with propagation vectors $\bm{q}$), including topologically non-trivial skyrmion lattices \cite{kurumaji2019skyrmion, khanh2020nanometric, hirschberger2019skyrmion}, have been uncovered in different quasi-2D lanthanide materials. It is believed that such phases originate from a combination of indirect exchange interactions mediated by the conduction electrons, some sense of magnetic frustration, and/or higher order interactions and anisotropy \cite{ozawa2016vortex, ozawa2017zero, hayami2021noncoplanar}. In addition to their ability to host real-space topological phases, such materials are exciting due to the potential for electronic structure engineering, where the multi-$q$ spin texture, or spin moir\'{e} superlattice (SMS), can modify the bandstructure and topology at the Fermi surface \cite{shimizu_spin_2021,kurumaji2025electronic}. Due to these exciting possibilities, it is therefore highly beneficial to explore the phase diagrams of members of this family of materials. The mapping of the phases, their structural determination, and connection to models, provides insight into their design principles.

\begin{figure*}[ht]
	\includegraphics[width=\textwidth]{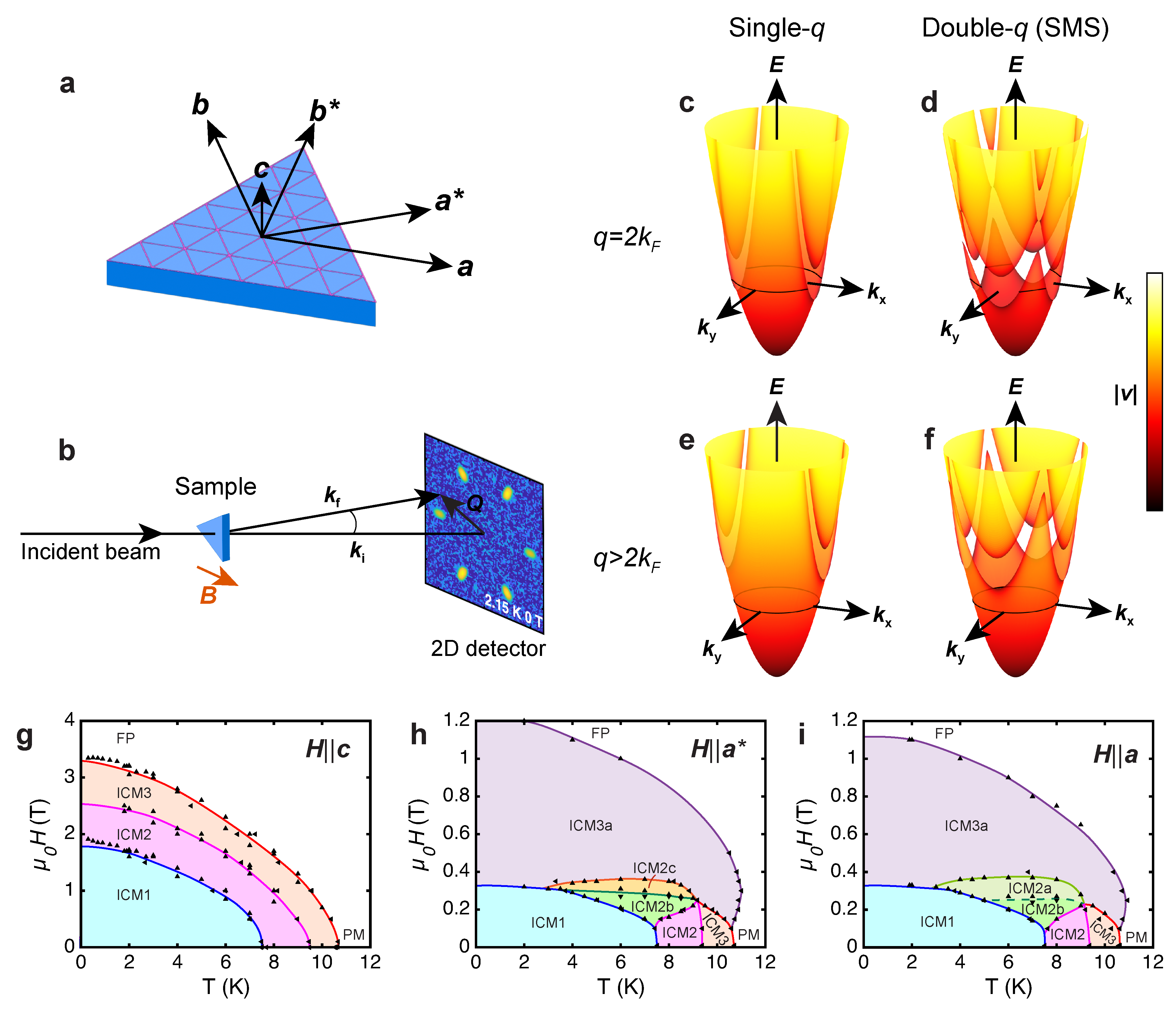}
	\caption{\label{fig:intro} \textbf{a} Schematic of the real and reciprocal lattice vectors of \ce{EuAg4Sb2} in relation to the rhombohedral morphology of the synthesized single crystals. The orientation of the Eu triangle lattice layers is indicated in pink. $a$, $b$, and $c$ are the lattice vectors for a hexagonal unit cell and $a^*$ and $b^*$ are the reciprocal lattice vectors. \textbf{b} Schematic of the SANS geometry. The sample is rocked about the horizontal and vertical axes to collect a 3D diffraction pattern. The magnetic field (orange) is transverse to the neutron beam and is rocked with the crystal. Two crystals in different orientations were used to measure the $H||a$ and $H||a^*$ phases.
    \textbf{c-f} Schematic 2D bandstructure of a parabolic electron band subject to a \textbf{c,e} single-$q$ or \textbf{d,f} double-$q$ (SMS) modulation with propagation vector \textbf{c,d} equal to ($q=2k_F$) or \textbf{e,f} slightly larger than ($q>2k_F$) the Fermi momentum. The color maps to the band velocity $|v|=\sqrt{v_x^2+v_y^2}=1/\hbar \nabla_{\bm{k}} E$.
    \textbf{g} The temperature-field phase diagram for $H||c$ (reproduced from \cite{kurumaji2025electronic}). PM: paramagnetic; FP: field-polarized; and ICM1-3: incommensurate magnetic modulation states. \textbf{h-i} The temperature-field phase diagram for $H||a^*$ and $H||a$, respectively. ICM2a-c and ICM3a are field-induced phases. Boundaries observed from upwards (downwards) field-sweeps are indicated with upwards (downwards) pointing triangles. Boundaries from field cooling are indicated with leftward pointing triangles. See supplementary section \phasesect\ for more information on the mapping of the in-plane phase diagrams.
 }
\end{figure*}

\ce{EuAg4Sb2} is a rhombohedral quasi-2D Eu triangle lattice semimetal which exhibits a rich phase diagram of incommensurate magnetic phases (ICMs) \cite{kurumaji2025electronic, green2025robust}. These phases are intimately connected with the electronic bandstructure of the material, where a $q=2k_F$ criterion is satisfied with the smallest cylindrical hole pocket, leading to a strongly renormalized bandstructure \cite{kurumaji2025electronic}.
%The combination of a electronically clean multi-$q$ spin texture where $q$ matches $2k_F$ to coherently renormalize the Fermi surface is the definition of a SMS.
The crystals grow with a platelet-like rhombohedral morphology, where the $a$ and $a^*$ directions can be readily identified by the crystal morphology (see Fig. \ref{fig:intro}a). The magnetic neutron diffraction data for these incommensurate textures can be readily obtained with small angle neutron scattering (SANS) experiments \cite{kurumaji2025electronic} (see Fig. \ref{fig:intro}b). The $H||c$ phase diagram has been explored previously, and a phenomenological anisotropic model with four-spin interactions was developed which predicts these phases \cite{kurumaji2025electronic}. At zero field or under $H||c$, three phases are accessible. the ground state ICM1 is a single-$q$ in-plane cycloid, while the higher temperature ICM2 and ICM3 phases are double-$q$ vortex lattices with different in-plane orientations \cite{neves2025polarized}. 
To resolve the anisotropy of the interactions in more detail, it is elucidating to explore the phase diagram with field in-plane. It is known that a host of additional phases can be accessed by rotating the magnetic field away from the $c$ axis \cite{green2025robust}, which is consistent with the phase boundaries identified in our own magnetization measurements (see Fig. \ref{fig:intro}g-i). ICM2 splits into three phases ICM2a/b/c (depending on in-plane field direction and strength) and ICM3 turns into ICM3a as field is rotated away from the $c$ axis. The ICM phases are summarized in Table \ref{tab:ICM}.

In this paper, we expand on the previous works by using SANS to identify the propagation vectors and multi-$q$ nature of multiple multi-$q$ SMS phases under the application of in-plane field. We show that the in-plane phases ICM2a, ICM2b, and ICM2c are rather unusual low symmetry multi-$q$ states, while ICM3a is a single-$q$ phase that can be seen as the logical evolution of either ICM3 or ICM1 with the application of in-plane field. We also compare the experimental phase diagram and phases with a phenomenological momentum-space model. Several additional nuanced evolutions of the intensity or location of magnetic diffraction peaks are discussed. Finally, we demonstrate a tight coupling between the propagation vector and multi-$q$ nature with the SMS transport behavior. The exploration of the phase diagram and modeling of this material provides additional insight into the design principles for the engineering and tunability of spin moir\'{e} materials with desirable band-structure features. Further, the ability to stabilize tunable multi-$q$ states through in-plane magnetic field separates this class of materials from conventional skyrmion lattice materials \cite{khanh2022zoology, hayami2025plane}.

\section{Results}
\subsection{In-Plane Phase Diagrams}
The phase diagram of \ce{EuAg4Sb2} can be mapped out with transport, magnetization, and heat capacity measurements \cite{kurumaji2025electronic, green2025robust}. The $H||c$ phase diagram is depicted in Fig. \ref{fig:intro}g, reproduced from \cite{kurumaji2025electronic}. ICM1, ICM2, and ICM3 are accessible at zero field below the transition temperature of 10.7 K. As field is turned on, the transition temperatures for all three phases decrease, reaching a field polarized saturation (FP) transition of 3.2 T at 2 K. Here, we have used magnetization measurements (see supplementary section \phasesect) to map out the $H||a^*$ and $H||a$ phase diagrams, depicted in Fig. \ref{fig:intro}h-i. Unlike the $H||c$ phase diagram, additional magnetic phases are accessible at finite field that are distinct from the zero field ICM1-3 phases. We follow the nomenclature of \cite{green2025robust}, which identified the in-plane phase boundaries but did not present any magnetic texture characterization, calling them ICM2a, ICM2b, ICM2c, and ICM3a, as each is an incommensurate magnetic modulation state that can be observed to originate from ICM2 or ICM3 as field is progressively rotated into the plane. The nuanced phase boundaries and finite temperatures required for ICM2a, ICM2b, and ICM2c hint at a complex competition of energies and thermal fluctuations to stabilize these different phases. Further, the similarity of the $H||a^*$ and $H||a$ phase diagrams suggests that the energy landscape in the plane is relatively isotropic. In the following, we probe each phase in more detail.

\subsection{ICM2a, ICM2b, ICM2c, and ICM3a}
\begin{figure*}[ht]
	\includegraphics[width=\textwidth]{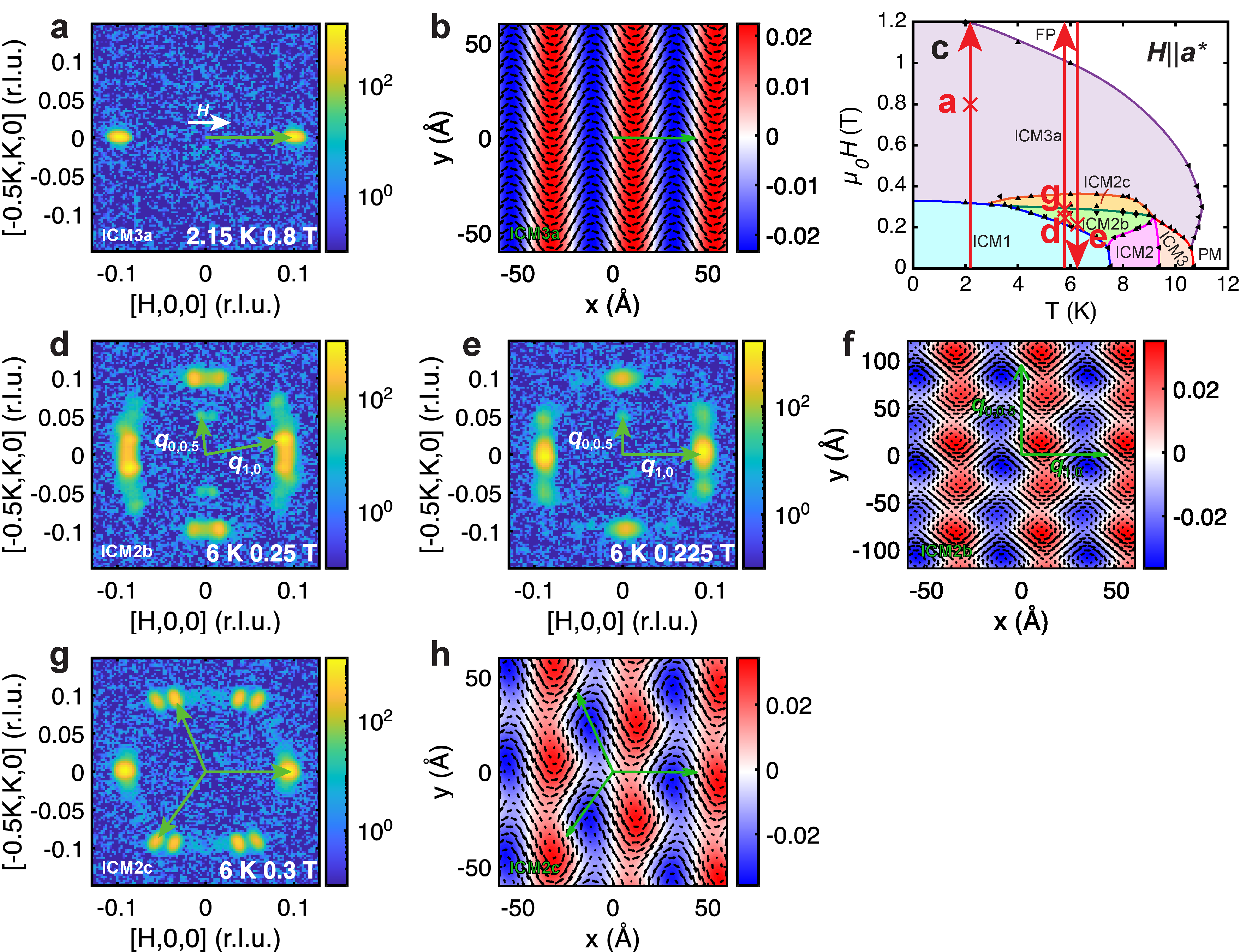}
	\caption{\label{fig:ICM2a-b} 
    Overview of representative in-plane-field SANS patterns and their corresponding real-space textures.
    \textbf{a} SANS diffraction pattern in sample S1 for ICM3a at 2.15 K with 0.8 T of field applied along the $a^*$ direction (horizontal axis). A set of two diffraction peaks corresponding to the single-$q$ propagation vector (see green arrow) are observed along the $a^*$ direction. \textbf{b} The corresponding simulated real space spin texture for ICM3a, with the colorscale showing the vorticity (see Eq. \ref{eq:vorticity}). Cartesian coordinates $x$ and $y$ are defined, where the $x$ axis is along the $a^*$ direction.
    \textbf{c} The in-plane phase diagram for magnetic $H||a^*$. The temperature and field conditions of \textbf{a,d,e,g} are indicated with red crosses.
    %\textbf{c-d} (\textbf{e-f}) Corresponding SANS diffraction pattern and simulated real space spin texture for ICM2b at 6 K and 0.25 T (ICM2c at 6 K and 0.3 T) with field sweeping up after zero field cooling.
    \textbf{d} The SANS diffraction pattern for ICM2b at 6 K and 0.25 T with field sweeping up after zero field cooling and 
    \textbf{e} after ramping field to 1.1 T, and then ramping field back down to 0.225 T. In both patterns, each double-$q$ domain is formed of two primary peaks at an approximate right angle with weaker half-order peaks. The propagation vectors are depicted with green arrows.
    \textbf{f} The corresponding real space spin texture for ICM2b corresponding to the diffraction pattern shown in \textbf{e}, with the colorscale showing the vorticity. \textbf{g} The SANS diffraction pattern for ICM2c at 0.3 T with field sweeping up after zero field cooling. The strongest diffraction peaks are along the $a^*$ direction, while eight additional diffraction peaks are present split about positions forming a hexagon with the primary peaks. \textbf{h} The corresponding real space spin texture for ICM2c corresponding to the diffraction pattern shown in \textbf{g}, with the colorscale showing the vorticity.
    All SANS data is inversion symmetrized, smoothed, $q_z$ integrated, and has a high temperature background subtracted.
 }
\end{figure*}

To uncover the nature of these phases, we examine their SANS diffraction patterns. We discuss three representative SANS patterns of ICM3a, ICM2b, and ICM2c (Fig. \ref{fig:ICM2a-b}(a,c,e), respectively), observed for $H\parallel a^*$. Additional detailed temperature and field dependent SANS measurements with $H||a$ and $H||a^*$ including ICM2a are included in supplementary section \moresans.

First, we discuss ICM3a, which consists of two peaks along the magnetic field direction. ICM3a can be entered directly from ICM1, ICM3, or ICM2b with the application of sufficiently strong in-plane field. ICM3a is single-$q$ with propagation vector (0.100,0,0.03) in reciprocal lattice units oriented along the field direction at 2.15 K, 0.8 T (see Fig. \ref{fig:ICM2a-b}a).
The single-$q$ nature is the same as the cycloidal ICM1 state \cite{green2025robust,neves2025polarized}
, while as there is a phase transition between ICM1 and ICM3a, the most natural transition is for the component of spin modulation along $q$ to vanish in ICM3a, as the field along $q$ makes the moment pointed anti-parallel to the field energetically unfavorable. This leaves a transverse spin modulation with a ferromagnetic component along the field direction (see Fig. \ref{fig:ICM2a-b}b, and Methods for additional details of the modeling). Though we cannot exclude the possibility of a proper screw spiral or conical phase, this phase is consistent with phenomenological modeling discussed below. This is sensible from an energetics perspective, as the spin is more aligned with the field than ICM1 or ICM3 as an intermediate to the field polarized state, and the local moment size is approximately the same on each site, satisfying the saturated moment condition. Note that in some regions near the low-field transition, we observe a weak second peak next to the primary peak (see Supplementary Fig. {\phaseevofig}a). This is likely a weak coexistence of ICM2b. When ICM3a is entered with $H||a$, the propagation vector follows the field direction (see SI Fig. \IIKa).

ICM2b can be entered from either ICM1 or ICM2 by applying in-plane field. ICM2b exhibits a rather unusually low symmetry SANS pattern which consists of two main peaks with weaker half-order peaks.
The two strong peaks are at $\bm{q}_{1,0}=$(0.079,0.018,-0.02) and $\bm{q}_{0,1}=$(-0.063,0.099,0), and with two weaker peaks at $\bm{q}_{0,0.5}=0.5\bm{q}_{0,1}$ and $\bm{q}_{1,0.5}=\bm{q}_{1,0}+0.5\bm{q}_{0,1}$ at 6 K, 0.25 T (see Fig. \ref{fig:ICM2a-b}\textbf{d}). A second copy of this pattern also exists as a mirror across the $b-c$ plane. This can be understood as a double-$q$ texture with principal propagation vectors $\bm{q}_{1,0}$ and $\bm{q}_{0,0.5}$, and the other vectors as higher harmonics, though the stronger $\bm{q}_{0,1}$ peak suggests the view that $\bm{q}_{0,0.5}$ and $\bm{q}_{1,0.5}$ are sub-harmonics is a more natural one. Similar behavior has been observed previously in other lanthanide metals \cite{sokolov2019metamagnetic, arachchige2024nanometric} and may be connected to higher order conduction-band mediated interactions. Additional theoretical insights on the stability of this phase are highly interesting. These propagation vectors are reminiscent of the low symmetry propagation vectors observed in ICM2, with the stronger peak along the field direction.

Assuming that each peak corresponds to a transverse spin modulation, and that the subharmonic peaks are in phase with the principal peaks, the resulting spin texture is visualized in Fig. \ref{fig:ICM2a-b}f. This phase may be seen as the intermediate breaking down of a vortex lattice phase. The vorticity, given by 
\begin{equation} \label{eq:vorticity}
\omega_z = (\nabla\times \bm{M})_z=\frac{\partial M_y}{\partial x}-\frac{\partial M_x}{\partial y},
\end{equation}
merges along the dimension perpendicular to the field, while still modulating in both directions. The sub-harmonics modulate the magnetism with a doubled unit cell along the vertical axis. This weak unit cell doubling is relatively uncommon, and may imply that the interactions in momentum space are highly frustrated so that the interaction at other wave vectors can contribute to the internal energy.

Of additional interest is the variety that this phase can display with different preparations. When entering ICM2b by lowering the field, the two rotated domains align along the horizontal and vertical directions (see Fig. \ref{fig:ICM2a-b}\textbf{e}). Additionally, the pattern will rotate to the applied in-plane field direction (see Supplementary Fig. {\VKa}-{\VIIIKa}). This ability to rotate a multi-$q$ spin texture so easily in-plane speaks to a remarkably flat energy landscape for the propagation vectors. This is consistent with the variety of propagation vectors that are observable in the zero field phases.%, as well as the ring of fluctuations observed above the transition temperature (unpublished data).%\cite{neves2025diffuse}.
The tunability of the texture orientation with in-plane field motivates further investigation of the impact on the bandstructure and transport properties.

Further, especially when entered through field cooling, at higher temperatures the subharmonic ICM2b peaks vanish (see Supplementary Figs. {\phaseevofig}c). Finally, for field applied along the $a$ direction, ICM2b begins as one domain aligned with the field. Upon increasing field, first the peaks perpendicular to field split, and then the peaks along the field split (see Supplementary Figs. \VKa-\VIIIKa). This transition was identified in \cite{green2025robust} as a transition from ICM2b to ICM2a. It is unclear if ICM2a and ICM2b represent distinct phases, or a continuous crossover.  These observations demonstrate the vast tunability of this phase with the application of magnetic field, and hint at a remarkably rich energy landscape. Additional work on understanding the origins and ramifications of this phase is of great interest.

ICM2c is entered from ICM2b by increasing the in-plane field. This phase exhibits a strong peak along the field direction with two additional sets of two peaks split along the field direction centered approximately about hexagonal positions about the $c$-axis. The strongest peak occurs at (0.093,0,-0.025) with weaker peaks at (-0.083,0.094,0.012) and (0.011,-0.094,0.012) (see Fig. \ref{fig:ICM2a-b}g). The side peaks are split laterally from a hexagonal position, and it is challenging to determine which peaks precisely belong to a single domain. We hypothesize that this phase is triple-$q$, though further experiments are required to confirm this. If all five propagation vectors are present in one domain, then this would be a quasicrystalline 5-q state as has been predicted previously \cite{solenov2012chirality}. This phase does not appear to be accessible with $H||a$, indicating that trigonal magnetocrystalline anisotropy of the material is relevant to the stability of this phase.

Assuming that the phase is triple-$q$ and the propagation vectors are transverse spin modulations like ICM2 and ICM3, the real-space phase is visualized in Fig. \ref{fig:ICM2a-b}h. Note that for a triple-$q$ state, there is a phase degree of freedom between the different propagation vectors. Various alternative possible textures are visualized in Supplementary Figure \phasechoice. Again, this phase can be understood as an intermediate phase between a double-$q$ vortex lattice and a single-$q$ spin modulation, where some modulation in both directions is visible, but the vorticity has merged along the direction perpendicular to the field. This conclusion is independent of exactly which $q$ vectors are chosen in a triple-$q$ state.

Finally, we note that under some conditions, especially at high temperatures, under field cooling, or at the phase boundary (see Supplementary Fig. {\phaseevofig}d, \IVKas, or \VIIIKas, \VIIIKdas, \CCCmTdas \ respectively), the upper and lower peaks become broader and merge, representing some disorder at the transitions and higher temperatures. Finally, we note that, unlike single-$q$ or double-$q$ phases, the triple-$q$ nature of this phase means that an additional degree of freedom, the relative phase of the spin modulations, is not just equivalent to a shift of origin \cite{shimizu2022phase}. We depict several choices of relative phase in Supplementary Fig. \phasechoice, though this does not impact the above conclusions.

% mention ICM4
% does this phase diagram obey gibbs rules? landau rules?

\subsection{Spin Moir\'{e} Superlattice}
Now we move to compare the connections of the propagation vectors measured \textit{via} SANS in the previous section to the electrical transport response in this material. In previous work, strong transport response has been found in some, but not all phases in \ce{EuAg4Sb2} \cite{kurumaji2025electronic,green2025robust}. This has been attributed to a commensuration condition where the propagation vector $q$ closely matches twice the Fermi momentum $2k_F$. When this occurs, a gap is opened in the bandstructure directly at the Fermi surface, maximally enhancing the mass of the bands and reducing the carrier density participating in electrical conduction. This is the electronic commensuration condition. A multi-$q$ texture further enhances this effect by gapping the Fermi surface more isotropically than a single-$q$ structure. See Fig. \ref{fig:intro}c-f for a schematic depiction of this process.

\begin{figure*}[ht]
	\includegraphics[width=\textwidth]{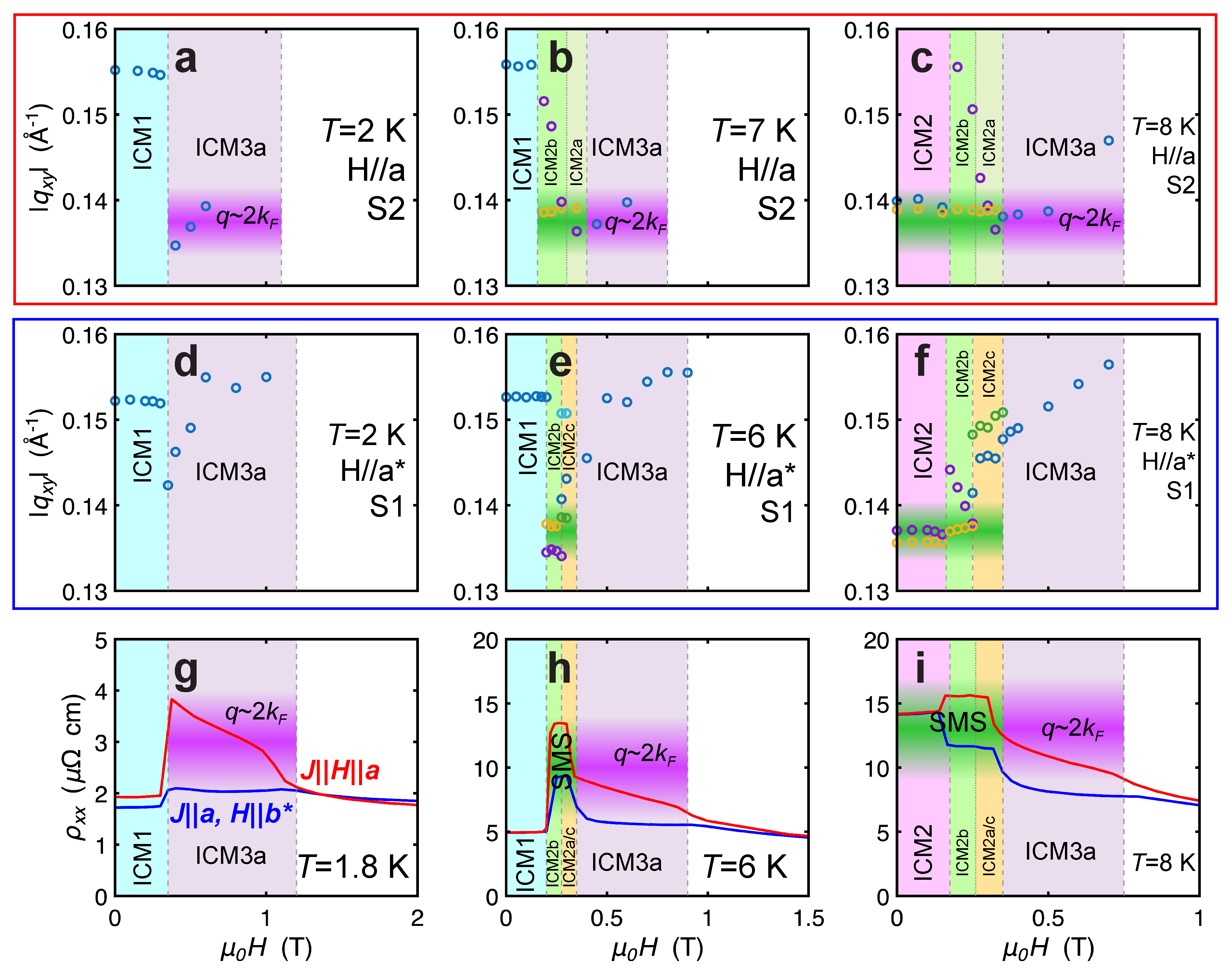}
	\caption{\label{fig:sms} \textbf{a-f} Magnitude of the in-plane component of the magnetic propagation vector as a function of in-plane field for magnetic field applied along the \textbf{a-c} $a$ and \textbf{d-f} $a^*$ directions measured in sample S2 and sample S1, respectively. The temperature of each field dependence is indicated within each panel. The magnetic phase is labeled and indicated with colored shading. The nature of the propagation vector is indicated with the color of the mark (see supplemental information for more information on the different propagation vectors). The electronic commensuration condition ($q=2k_F$) is highlighted with a purple haze. Multi-$q$ states (the SMS condition) are highlighted with a green haze. \textbf{g-i} The longitudinal resistivity along the $a$ axis for field applied along $a$ (red) and $b^*$ (blue, perpendicular to the $a$ axis and crystallographically equivalent to $a^*$) at 1.8 K, 6 K, and 8 K, respectively. The resistivity enhancement is emphasized with a purple/green haze.
 }
\end{figure*}

To test this, we compare the in-plane component of $q$, $|q_{xy}|=\sqrt{q_x^2+q_y^2}$ (which is close to the full length of $q$ as the out-of-plane component is much smaller) to the longitudinal transport response. The $q_{xy}$ for $H||a$ and $H||a^*$ for various temperatures is shown in Fig. \ref{fig:sms}\textbf{a-f}. The longitudinal response at three temperatures is depicted in Fig. \ref{fig:sms}\textbf{g-h} for current is applied along the $a$ axis, while field is applied along either the $a$ (red) or $b^*$ (blue) direction. The magnitude of $|q_{xy}|$ for $H||a$ and $H||a^*$ varies slightly when compared at zero field. This is likely due to variation between crystals, as two crystals from different batches were used for the different orientations. This may highlight the sensitivity of $q$ to the Fermi surface, and motivates future careful doping dependence studies.

We note that in all cases, a $|q_{xy}|$ near the value of $\sim$ 0.1375 \AA$^{-1}$ is correlated with an increase in resistivity above the field-polarized value, as expected for a more strongly gapped Fermi surface. Further, the strongest response is observed in ICM2, and ICM2a-c which are all multi-$q$ states. This highlights how the multi-$q$ nature can work together with the $q\sim2k_F$ condition to create a stronger transport response in the SMS state (see Fig. \ref{fig:intro}c-f). The single-$q$ ICM3a has a strong transport response only when $q$ is aligned along the current direction for $H\parallel a$ (red curve in Fig. \ref{fig:sms}g) when the $|q_{xy}|$ is close to the electronic commensuration condition. This is consistent with the above scenario as the gap opening happens at $\bm{k}$ points of the electron bands with the velocity $\bm{v}_{F}$ ($=\frac{\partial E}{\hbar \partial \bm{k}}$) along the current direction (Fig. \ref{fig:intro}c). Additionally, $q$ is less well matched to $2k_F$ in ICM3a when field is applied along the $a^*$ direction, amplifying this effect. These findings are summarized in Table \ref{tab:ICM}. Additional $q$ vector dependence and transport response data can be found in supplementary materials.

\begin{table*}[t]
\caption[ICM Phase Summary]{Summary of the ICM phases; whether a resistivity jump is observed, if they are multi-$q$ (SMS) or single-$q$, if they satisfy the $q=2k_F$ condition, and a brief description of the phase.}\label{tab:ICM}%
\centering{%
\begin{tabular*}{1.0\textwidth}{@{\hspace*{1.5em}}@{\extracolsep{\fill}}ccccc@{\hspace*{1.5em}}}
\\[-0.5em]
\toprule
Phase & Resistivity Jump & Multi-$q$ (SMS) or Single-$q$ & $q=2k_F$ & Description\\
 \midrule
 ICM1 & No & Single & No & Cycloid\\
 ICM2 & Yes & Double & Yes & Vortex lattice\\
 ICM3 & Yes & Anisotropic Double & Partly Yes & Vortex lattice\\
 ICM2a,b & Yes & Anisotropic Double & Partly Yes & Intermediate vortex phase\\
 ICM2c & Yes & Anisotropic Double/Triple & Partly Yes & Intermediate vortex phase\\
 ICM3a & Partly (when $H||J||a$) & Single & At low field & Transverse modulation phase\\
\bottomrule
\end{tabular*}
}%
\end{table*}

\subsection{Phenomenological Modeling}
\begin{figure*}[ht]
	\includegraphics[width=\textwidth]{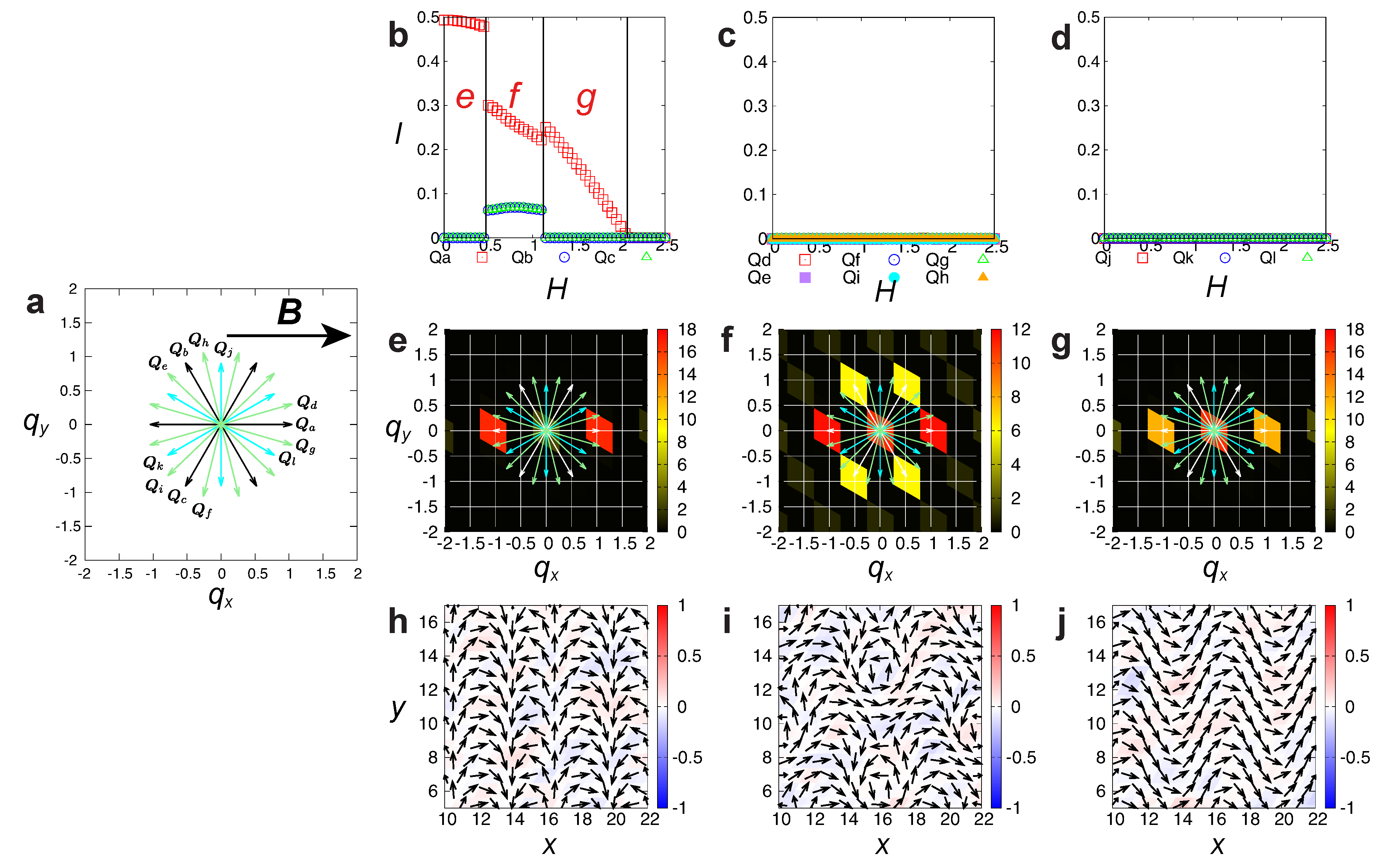}
	\caption{\label{fig:hayami_x} Model phase diagram for field applied along the $a$ direction. \textbf{a} Propagation vectors considered in model. Black arrows correspond to the ICM1 (and ICM3) vectors, green arrows correspond to the ICM2 vectors, and blue arrows correspond to the ICM3 vectors. $q_x$ and $q_y$ are Cartesian coordinates in the reciprocal space, where $q_y$ is along the $a^*$ direction. \textbf{b-d} The model peak intensity for each of the peaks as a function of applied magnetic field. \textbf{e-g} Real-space model spin structure in several phases for increasing field. The in-plane spin is depicted with an arrow, and the $z$ component is indicated with red.
 }
\end{figure*}

\begin{figure*}[ht]
	\includegraphics[width=\textwidth]{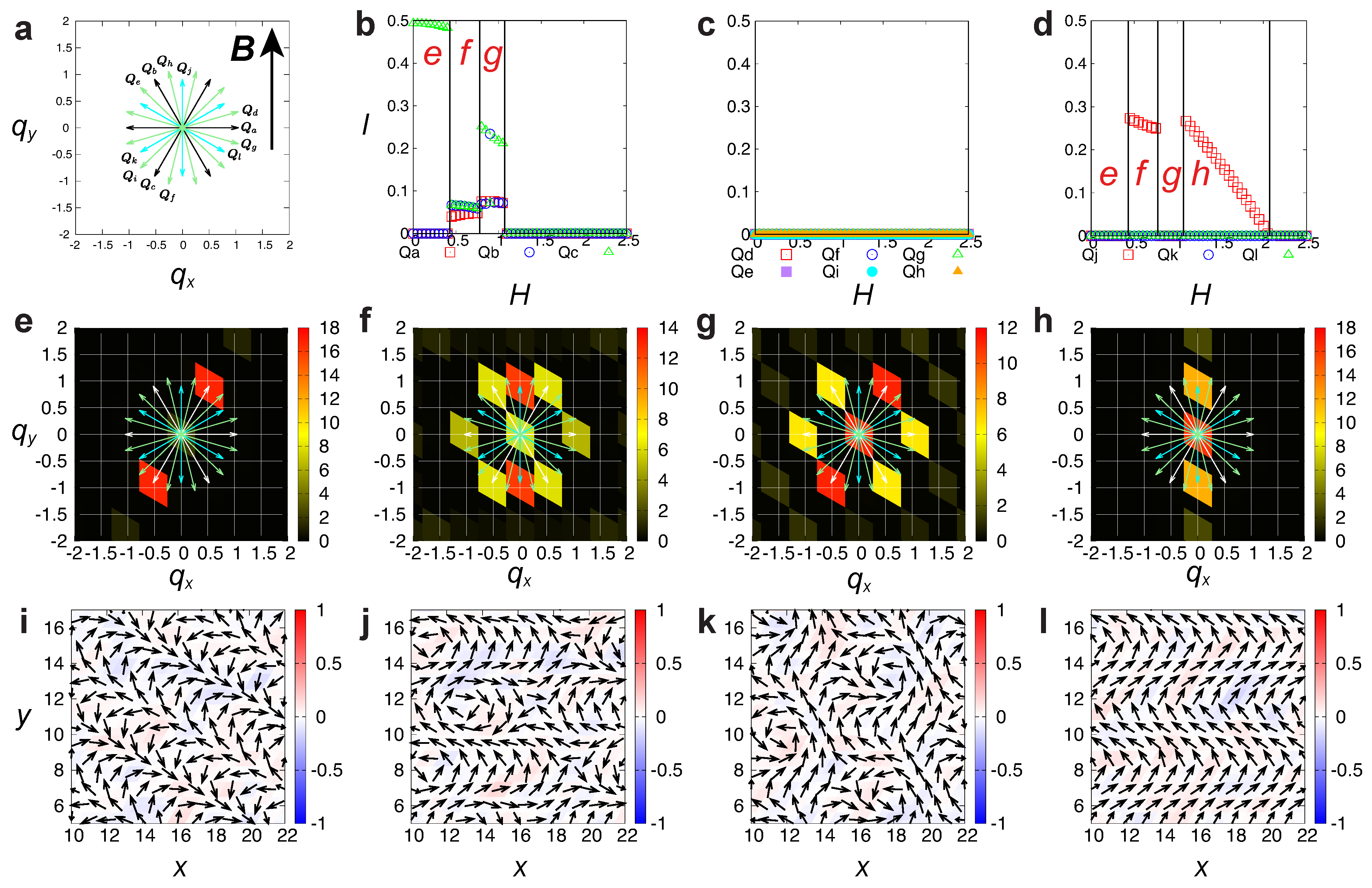}
	\caption{\label{fig:hayami_y} Model phase diagram for field applied along the $a^*$ direction. \textbf{a} Propagation vectors considered in model. Black arrows correspond to the ICM1 (and ICM3) vectors, green arrows correspond to the ICM2 vectors, and blue arrows correspond to the ICM3 vectors. \textbf{b-d} The model peak intensity for each of the peaks as a function of applied magnetic field. \textbf{e-h} Real-space model spin structure in several phases for increasing field. The in-plane spin is depicted with an arrow, and the $z$ component is indicated with red. $x$ and $y$ are Cartesian coordinates in the $ab$ plane, where $x$ is along the $a$ axis.
 }
\end{figure*}

To further examine the energetic origins of these phases, we now turn to phenomenological modeling. Using the spin model obtained previously \cite{neves2025polarized}, we calculate the phase diagram for $H||a$ (Fig. \ref{fig:hayami_x}) and $H||a^*$ (Fig. \ref{fig:hayami_y}) directions as a function of field. This model accurately reproduces both ICM1 and ICM3a as the low and high field incommensurate ground states for both field directions. For $H||a$, this model predicts one intermediate-field phase which bears some resemblance to ICM2a-c. For $H||a^*$, this model predicts two intermediate phases, one which is similar to the intermediate $H\parallel a$-model phase, and a second one. Naively, these intermediate multi-$q$ phases may be attributed to the bond-dependent anisotropic interaction. Since this interaction favors a fan-type oscillation, there is an energy cost when the magnitude of the spins is fixed. Such an effect can become smaller when the magnetic field is larger, which results in the single-$q$ ICM3a. While these intermediate phases are not present at zero temperature in \ce{EuAg4Sb2}, related phases (ICM2a-c) are present at finite temperature. This suggests that this model is reasonably relevant for the actual \ce{EuAg4Sb2} magnetism, though more work is required to accurately account for the complete temperature-dependent phase diagram.
% In my notation, I set the translational vectors of the triangular lattice as a1=(1,0,0) and a2=(-1/2, \sqrt{3}/2, 0); x means the direction of a1.
% x = a

\section{Discussion}
In this paper we have mapped and reported the phase diagram along the high symmetry in-plane directions of \ce{EuAg4Sb2}. We have also uncovered the magnetic propagation vectors and their evolution in each of the phases stabilized with in-plane field. We connect the propagation vectors with the transport response, noting a strong coupling between multi-$q$ SMS behavior, $q\sim2k_F$, and the transport response. Further, we compare the observed phases with those anticipated from a phenomenological model. Together, these observations reveal a highly tunable and nuanced energy landscape which can seed numerous single- and multi-\textit{q} incommensurate spin textures.

Several features observed in this work warrant further study. The broad peaks observed in some phases under some preparations (and ability for some peaks to evolve greatly with field) motivates future studies of the characteristics of the disorder or domain wall physics with \textit{e.g.} tomographic \cite{henderson2023three} or real-space methods \cite{wolf2022unveiling, seki2022direct}. Of additional interest is to study how these phases evolve as field is rotated from in-plane to out-of-plane \cite{green2025robust} or under chemical tuning, where the coupling of the magnetic modulations and the Fermi surface may evolve. Finally, it is worthwhile to confirm the real-space models presented here with polarized neutron scattering or real-space probes, as well as the multi-\textit{q} domains with domain selecting techniques as were employed in previously to solve the out-of-plane phases \cite{neves2025polarized}.

Overall, the coupling of these new tunable SMSs with the electronic structure and properties is of significant future interest for any potential spintronic or other technological applications. More generally, these phases also represent a rare example of (several) multi-\textit{q} incommensurate phase(s) stabilized with the application of in-plane field \cite{khanh2022zoology, hayami2025plane}. Unlike kagome metals or conventional skyrmion hosts, where typically only a single multi-q phase pocket exists in the phase diagram, EuAg4Sb2 uniquely exhibits its cascade of several distinct multi-$q$ phases under in-plane field, underscoring its unusually rich frustrated energy landscape. Further, this work demonstrates that a close matching of $q$ and $2k_F$ for a cylindrical pocket in a triangle lattice material strongly affects the transport response, and multi-$q$ textures are more effective at this gap-opening behavior. These insights into the design of tunable electronic properties may be of broad potential relevance to other frustrated magnet and SMS systems.

% says something about anisotropy of magnetic interactions

\section{Methods}
\subsection{Synthesis and Characterization}
Single crystal samples were synthesized from a published self flux method \cite{kurumaji2025electronic}. Phase identity was confirmed with powder x-ray diffraction measurements. Magnetization measurements were performed with a Quantum Design Magnetic Property Measurement System.

\subsection{Transport}
Electrical transport measurements were performed by a conventional five-probe method with a typical AC excitation current of 1 mA at typical frequency near 15 Hz.
The transport response in low temperatures and a magnetic field was measured using a commercial superconducting magnet and cryostat.
The obtained longitudinal and transverse signals were field symmetrized and antisymmetrized, respectively, between field-increasing and decreasing scans to correct for contact misalignment.

\subsection{SANS}
SANS measurements were made at the D33 instrument at the Institut Laue-Langevin using 4.7 \AA neutrons and a high temperature superconducting magnet. Rocking curves were performed over the horizontal and vertical axes to sample the three dimensional diffraction patterns. Data was analyzed with the GRIP \cite{neves2024grasp} module of GRASP \cite{dewhurst2023graphical}. Data had a high temperature background subtracted to isolate the magnetic signal.

\subsection{Phase Modeling}
The magnetic phases present in \ce{EuAg4Sb2}) can be represented with the equation 
\begin{equation} \label{eq:1}
    \bm{M}(\bm{r})=\text{Re}\left[\sum_i \bm{M}_i \exp \left( i \bm{q}_i \cdot \bm{r} \right)\right],
\end{equation}
where $i$ indexes the magnetic propagation vector(s) $\bm{q}_i$, $\bm{r}$ is the real-space position, $\bm{M}(\bm{r})$ is the real-space magnetization density, and the $i^{th}$ spin wave has complex Fourier amplitude $\bm{M}_i$. Uniform magnetization is accommodated as a $\bm{q}=0$ term. Based on the tendency of the system to prefer $ab$-plane transverse spin modulations except for the $ab$-cycloidal ICM1 \cite{neves2025polarized}, a transverse spin modulation is assumed for all phases except for ICM1. This should be confirmed with future measurements of nuclear satellite peaks with polarized neutrons under in-plane fields. The relative magnitude of the modulation for each propagation vector is obtained by taking the square root of the scattering intensity.
% should normalize by Eu form factor actually
For multi-$q$ phases with three or more non-zero propagation vectors (single-$q$ and double-$q$ phases can wrap this into a gauge freedom that represents a spatial translation), there is some ambiguity in determining the relative phase between the different propagation vectors \cite{shimizu2022phase}. The phase relation may be in some cases determined through symmetry considerations, however in other cases microscopic probes may be necessary. For ICM2b, several potential phase factors are plotted in Supplementary Plot \phasechoice.
% how to decide? which fulfills the saturation moment condition the best? no they are all the same
% include other phase examples in SI
The $\bm{q}=0$ uniform magnetization component in finite field was determined from in-plane magnetization measurements. The strongest peak in each phase is assumed to have a spin component with the full saturation moment of Eu$^{2+}$. This too could be refined with polarized neutron measurements of the integer and satellite peaks.
% how to calibrate to SANS intensity? use ICM1 in zero field probably

The unpolarized neutron scattering cross section is
\begin{equation}
\label{eq:sigmaM}
    \frac{d\sigma}{d\Omega} = \left( \frac{\gamma r_0}{2\mu_\text{B}}\right)^2 \left| \left<\bm{M}_\perp(\bm{Q}) \right> \right|^2
\end{equation}
where $\gamma=1.913$, $r_0=\mu_0e^2/(4\pi m_e)=2.818\times10^{-15}$ m, $\mu_\text{B}=9.274\times10^{-24}$ J/T is the Bohr magneton, $\bm{Q}$ is the neutron momentum transfer, and $\left<\bm{M}_\perp(\bm{Q}) \right>$ is the expectation value of $\bm{M}_\perp(\bm{Q})=\hat{\bm{Q}}\times\{\bm{M}(\bm{Q})\times\hat{\bm{Q}}\}$, the component of $\bm{M}(\bm{Q})$ perpendicular to $\bm{Q}$, with $\hat{\bm{Q}}$ being the normalized unit vector along $\bm{Q}$. $\bm{M}(\bm{Q})$ is the Fourier transform of the real-space spin modulation $\bm{M}(\bm{r})$:
\begin{equation}
    \bm{M}(\bm{Q})=\int \bm{M}(\bm{r}) \exp (-i\bm{Q} \cdot \bm{r}) d^3 \bm{r}.
\end{equation}
Hence, magnetic neutron diffraction directly measures the square of the Fourier transform of the real-space magnetization density orthogonal to $\bm{Q}$.

\subsection{Phenomenological Modeling}
For the phenomenological model used to calculate the in-plane magnetic field dependent phase diagram, a momentum-space Hamiltonian \cite{hayami2024stabilization} is constructed following the model developed for the out-of-plane phase diagram \cite{neves2025polarized}
as
\begin{equation}
    H = -2 J \sum_{\nu,\alpha,\beta} \Gamma^{\alpha \beta}_\nu S^\alpha_{\bm{Q}_\nu} S^\beta_{-\bm{Q}_\nu} - H_x \sum_i S^x_i - H_y \sum_i S^y_i + H_{4}
\end{equation}
where $J$ is an overall energy scale, $\Gamma$ is a generalized anisotropic spin interaction in momentum space, $S$ is the Fourier transformed spin moment, $\alpha$ and $\beta$ are the cartesian directions, $\nu$ is the index of the propagation vector (see included propagation vectors in Fig. \ref{fig:hayami_x}a), and $i$ is a site index. 
The second term in the Hamiltonian is the Zeeman energy term. $H_4$ is a four-spin interaction term of the form
\begin{equation}
    H_4 = - \frac{B}{N} (\bm{S}_{\bm{Q}_{e}} \cdot \bm{S}_{-\bm{Q}_{e}}) (\bm{S}_{\bm{Q}_{i}} \cdot \bm{S}_{-\bm{Q}_{i}})
\end{equation}
where $\bm{Q}_{e}$ and $\bm{Q}_{i}$ are the two propagation vectors of ICM2, and N represents the total number of spins. We also consider the symmetry-related four-spin interactions.
$\Gamma_{\bm{Q}}$ is given as
\begin{equation} \label{eq:gamma_q}
    \Gamma_{\bm{Q}} = \begin{pmatrix}
        G+A\cos(2\theta) & -A\sin(2\theta) & 0\\
        -A\sin(2\theta) & G-A\cos(2\theta) & 0\\
        0 & 0 & 0\\
        \end{pmatrix}
\end{equation}
where $\theta$ is the angle made between $\bm{q}$ and the $x$ axis. The value of $\Gamma_{\bm{Q}}$ is in general a continuous function of $\bm{q}$, but we only consider it at the discrete $\bm{q}$ vectors approximately associated with ICM1-3 in this model, as depicted in Fig. \ref{fig:hayami_x}a, and the $B$ term is only included connecting the ICM2 vectors.

For the momentum-resolved interaction in Eq.\ref{eq:gamma_q}, we adopt the following wave vectors following \cite{neves2025polarized}: 
$\bm{Q}_a= Q(1,0)$, $\bm{Q}_b= Q(-1/2,\sqrt{3}/2)$, $\bm{Q}_c= Q(-1/2,-\sqrt{3}/2)$, $\bm{Q}_d=(Q, Q')$, $\bm{Q}_e=(-Q/2-\sqrt{3}Q'/2, \sqrt{3}Q/2-Q'/2)$, $\bm{Q}_f=(-Q/2+\sqrt{3}Q'/2, -\sqrt{3}Q/2-Q'/2)$, $\bm{Q}_g=(Q, -Q')$, $\bm{Q}_h=(-Q/2+\sqrt{3}Q'/2, \sqrt{3}Q/2+Q'/2)$, $\bm{Q}_i=(-Q/2-\sqrt{3}Q'/2, -\sqrt{3}Q/2+Q'/2)$, $\bm{Q}_j= Q''(0, 1)$, $\bm{Q}_k= Q''(-\sqrt{3}/2, -1/2)$, and $\bm{Q}_l= Q''(\sqrt{3}/2, -1/2)$ with $Q=\pi/3$, $Q'=\sqrt{3}\pi/18$, and $Q''=\sqrt{3}\pi/6$, where we consider the interaction parameters $G_1=1$ and $A_1=0.05$ for $(\bm{Q}_a, \bm{Q}_b, \bm{Q}_c)$, $G_2=0.95$ and $A_2=0.1$ for $(\bm{Q}_d, \bm{Q}_e, \bm{Q}_f, \bm{Q}_g, \bm{Q}_h, \bm{Q}_i)$, and $G_3=0.9$ and $A_3=0.16$ for $(\bm{Q}_j, \bm{Q}_k, \bm{Q}_l)$. 

The optimized spin configurations for the model are obtained through an iterative simulated annealing scheme combined with single-spin updates based on the Metropolis Monte Carlo algorithm. Periodic boundary conditions were imposed throughout, and simulations were performed for multiple lattice sizes to verify the robustness and convergence of the numerical procedure.

\subsection{Acknowledgments}
This work was funded, in part, by the Gordon and Betty Moore Foundation EPiQS Initiative, grant no. GBMF9070 to J.G.C. (instrumentation development, and DFT calculations); the US Department of Energy (DOE) Office of Science, Basic Energy Sciences, under award no. DE-SC0022028 (material development); the Office of Naval Research (ONR) under award no. N00014-21-1-2591 (advanced characterization); and the Air Force Office of Scientific Research (AFOSR) under award no. FA9550-22-1-0432 (magnetic structure analysis). S. Hayami was supported by JSPS KAKENHI (JP23H04869), JST CREST (JPMJCR23O4), and JST FOREST (JPMJFR2366).
ILL D33 neutron scattering experiments may be accessed at https://doi.ill.fr/10.5291/ILL-DATA.5-71-2 and hhtps://doi.ill.fr/10.5291/ILL-DATA.5-71-22.
J.S.W. acknowledges financial support from the Laboratory for Neutron Scattering and Imaging at PSI for an extended research visit of P.M.N., as well as funding from the Swiss National Science Foundation (SNF) under grant no. 200021$\textunderscore$188707. This work is based partly on experiments performed at the Swiss spallation neutron source SINQ, Paul Scherrer Institute, Villigen, Switzerland.

\subsection{Contributions}
Small angle neutron scattering was performed by P.M.N. and J.P.W. with support from J.S.W. and R.C.. T.K. and C.I.J.I. synthesized single crystals for SANS experiments. T.K. characterized the phase diagrams, and measured transport properties. P.M.N. analyzed the data and constructed the real-space diagrams. S. H. constructed the phenomenological model. All authors contributed to writing the manuscript. J.G.C. coordinated the project.

\subsection{Supporting Information}
Magnetization data used to construct in-plane phase diagrams, additional transport measurements, analysis of the zero-field phases with SANS, additional notes on the evolution of the phases, parameterization of the propagation vectors, a note on the phase of the vectors in the ICM2b state, and additional SANS measurements are provided in the supporting information.

% \vfill

% \pagebreak

% \clearpage

% \bibliography{references}

\end{document}